\documentclass[a4paper,fleqn,usenatbib]{mnras}

\usepackage[T1]{fontenc}
\usepackage{ae,aecompl}
\usepackage[dvipdfmx]{graphicx}

\usepackage{graphicx}	
\usepackage{amsmath}	
\usepackage{amssymb}	
\usepackage{subfig}
\usepackage{adjustbox}
\usepackage{booktabs}

\title[X-ray flaring activity of 1ES\,1959+650]{\textit{AstroSat} observation of the HBL\,1ES\,1959+650 during its October 2017 flaring}

\author[Shah et. al.]{
Zahir Shah$^{1}$\thanks{E-mail: zahir@iucaa.in}, Savithri H. Ezhikode$^{1}$\thanks{savithri@iucaa.in}, Ranjeev Misra$^{1}$ and Rajalakshmi T. R.$^{2}$ 
\\
$^{1}$Inter-University Center for Astronomy and Astrophysics, Post Bag 4, Ganeshkhind, Pune, India - 411007\\
$^{2}$ School of Pure \& Applied Physics, Mahatma Gandhi University, Kottayam, Kerala, India - 686560
}

\date{Accepted XXX. Received YYY; in original form ZZZ}

\pubyear{2021}

\begin{document}
\label{firstpage}
\pagerange{\pageref{firstpage}--\pageref{lastpage}}
\maketitle

\begin{abstract}
  We present the results of the X-ray flaring activity of 1ES\,1959+650 during October 25-26, 2017 using \textit{AstroSat} observations.  The source was  variable in the  X-ray. We investigated the evolution of the  X-ray spectral properties of the source by dividing the total observation  period ($\sim 130$ ksecs) into time segments of 5 ksecs, and fitting the SXT and LAXPC spectra for each segment. Synchrotron emission of a broken power-law particle density model provided a better fit than the log-parabola one. The X-ray flux and the normalised particle density at an energy less than the break one, were found to anti-correlate with the index before the break. However, a stronger correlation between the density and index was obtained when a delay of $\sim 60$ ksec was introduced. The amplitude of the normalised particle density variation $|\Delta n_\gamma/n_\gamma| \sim 0.1$ was found to be less than that of the index $\Delta \Gamma \sim 0.5$.
  We model the amplitudes and the time delay in a scenario where the particle acceleration time-scale varies on a time-scale comparable to itself. In this framework, the rest frame acceleration time-scale is estimated to be $\sim 1.97\times10^{5}$ secs and the emission region size to be $\sim 6.73\times 10^{15}$ cms.
\end{abstract}



\section{Introduction}

 Blazars are radio-loud Active Galactic Nuclei (AGN) with powerful relativistic jet pointing close to the line of sight of observer \citep{1995PASP..107..803U}. The small inclination of relativistic jet  results in Doppler boosted emission and other extreme properties from blazars like rapid variability across the electromagnetic spectrum \citep{2003ApJ...596..847B} and non-thermal emission extending from radio to GeV/TeV energies \citep{1997ARA&A..35..445U}. 
The spectral energy distribution (SED) of blazars has two prominent broad components: low energy component which peaks at optical/UV/X-ray energies and high energy component which peaks at MeV/GeV energies. According to leptonic model, two different emission processes are responsible for this broadband emission viz. the synchrotron process in which highly relativistic leptons in the presence of  magnetic field produces radio to UV/X-ray energies \citep{1981Natur.293..714B} and the inverse-Compton (IC) process in which low energy seed photons gets up-scattered to high energies giving rise to X-ray/GeV/TeV spectrum. The seed photons for the IC scattering can be synchrotron photons produced by the same population of leptons (synchrotron self Compton (SSC):  \citealp{1974ApJ...188..353J}, \citealp{1992ApJ...397L...5M}, \citealp{1993ApJ...407...65G}) or  photons  entering  external to jet  (external Compton (EC): \citealp{1992A&A...256L..27D}, \citealp{1994ApJ...421..153S}, \citealp{2000ApJ...545..107B}, \citealp{2017MNRAS.470.3283S}).  On the other hand, the high energy emission in blazars can be also produced by  relativistic protons via hadronic processes,  such as the proton synchrotron process and pion production \citep{1992A&A...253L..21M,2001APh....15..121M}.
Based on the characteristics of their optical spectrum, blazars are broadly classified as BL\,Lac objects and Flat Spectrum Radio Quasars (FSRQ). They are also classified based on the peak frequency of the synchrotron component: Low energy peaked BL Lac objects LBLs; $\nu_{p,{syn}}<10^{14}$ Hz ), Intermediate energy peaked BL Lac objects (IBLs; $10^{14}Hz<\nu_{p_{syn}}<10^{15}$ Hz) and High energy peaked BL Lac objects (HBLs; $\nu_{p,{syn}}>10^{15}$ Hz)\citep{2010ApJ...716...30A}. In case of HBL sources, single population of relativistic electrons generally produces the broadband emission in such a way that radio to X-ray emission is attributed to synchrotron process while hard X-ray to $\gamma$-ray emission is attributed to SSC process \citep{1997A&A...320...19M}.   HBLs are generally less variable in optical band than LBL sources \citep{2014JApA...35..241G, 2014MNRAS.439..690H}, however during the flaring their variability increases by significant amount.

1ES\,1959+650 is an HBL source located at redshift z=0.048
\citep{1996ApJS..104..251P}. This source was first observed in radio
band using NRAO Green Bank Telescope \citep{1991ApJS...75.1011G}, and
later the X-ray emission from it was detected during the Einstein IPC
Slew Survey \citep{1992ApJS...80..257E}. 
The VHE  emission from
1ES\,1959+650  was first detected by the Seven Telescope Array group
in 1999 \citep{1999ICRC....3..370N} and it became among the first
extra-galactic sources detected at VHE energies
\citep{2003ApJ...583L...9H}.  
Numerous exceptional flaring events has been reported from
1ES\,1959+650 across the electromagnetic spectrum (EMS) including the intense activity at very high energies (GeV/TeV). In
addition to broad emission like features, 1ES\,1959+650  also exhibits
orphan TeV flares \citep{2004ApJ...601..151K, 2014ApJ...797...89A}. During the simultaneous multi-wavelength
flaring, the X-ray and $\gamma$-ray fluxes of 1ES\,1959+650 are well
correlated, which is is usually explained in-terms of one-zone
Synchrotron Self-Compton (SSC) models. However, orphan VHE flare
challenges such interpretation and instead models like hadronic synchrotron mirror
model are incorporated to explain such flares \citep{2005ApJ...621..176B}. 
Further, the uncorrelated
multi-wavelength emission can be also explained by multi-zone SSC or
\citep{2008ApJ...689...68G} or EC process \citep{2004ApJ...601..151K}.

The X-ray spectrum of 1ES\,1959+650 varies significantly during
  the active states. In this source, the correlation between the X-ray
  flux and the spectral shapes have been studied in several works
  \citep[e.g. ][]{2002ApJ...571..763G, 2006ApJ...644..742G,
    2016MNRAS.457..704K, 2018MNRAS.473.2542K, 2018A&A...611A..44P, 2020A&A...638A..14M}.
  These correlation study showed that the spectral hardening increases
  as the flux increases i.e., {\it harder-when brighter} behaviour is
  observed, a general feature found in blazar
  \citep[e.g. ][]{1997ARA&A..35..445U, 2006ApJ...646...61G,
    2015ApJ...807...79H,
    2017ApJ...848..103K}. \citet{2006ApJ...644..742G} showed that {\it
    harder-when brighter} behaviour in flares on time scales of days
  is caused by a variation of the Doppler factor, the magnetic field,
  and/or break energy, whereas the shift of the maximum energy of
  accelerated electrons drives the longer timescales (months time
  scale) correlation between flux and spectral index. The {\it harder
    when brighter} behaviour on longer timescales is additionally
  accompanied by higher variability amplitude at harder energies and
  vice versa \citep{2006ApJ...646...61G}. This behaviour is also
  associated with the shift in the synchrotron peak of the SED
  \citep{2016MNRAS.457..704K, 2018ApJS..238...13K}. The synchrotron
  peak shift towards higher energy as the flux increases and vice
  versa.  Further, {\it harder when brighter} spectrum together with
  the hysteresis loop in the hardness ratio--flux plane provides
  important information about the acceleration and cooling process:
  the clockwise pattern indicates that the acceleration timescales are
  smaller than the cooling timescale, whereas anticlockwise pattern is
  expected when acceleration and cooling timescales are same
  \citep{1999APh....11...45K, 2018ApJS..238...13K}.

Several enhanced flux states has been reported from 1ES\,1959+650 over
broadband energy, during the period 2015--2016 (e.g.,
\cite{2016MNRAS.461L..26K}. During this period, 1ES\,1959+650 had
shown prolonged X-ray activity with powerful X-ray emission, it
reached in a state of record count rate ($> 20~counts~s^{-1}$) in the
energy range of 0.3--10~KeV \citep{2016MNRAS.461L..26K}. This outburst
placed 1ES\,1959+650 to the list of only three blazars detected with
X-ray count rate exceeding 20~counts~s$^{-1}$. Another
  X-ray flaring activity was observed in the source
  \citep{2017ATel10743....1K} since 2017 June.  The Neil Gehrels Swift Observatory (here after \textit{Swift}) X-ray telescope (XRT)
  observations reported a count rate (0.3--10~keV) of
  $\sim$~39~counts~s$^{-1}$ in September 2017, much higher than the
  value reported in the previous year \citep{2018ApJS..238...13K}. A
  month later, a Target of Opportunity observation of the source was
  triggered with \textit{AstroSat}. The observation showed a 0.7--30~keV flux
  of $\sim{\rm 1.4\times10^{-9}erg~cm^{-2}s^{-1}}$ similar to the
  simultaneous \textit{Swift} observation taken on 2017 October 25.

In this work, we report the X-ray and UV analysis of
\textit{AstroSat} observations  of 1ES~1959+650 during October 2017.
The high flux rate and rapid variability in the X-ray band
allowed for time-resolved
spectral analysis which revealed  correlated spectral parameter variations
on time-scales of hours. We interpret the results by comparing with
predictions obtained from solving the linearized kinetic equation
for particle distribution having sinusoidal perturbations.
In the next section, we provide details of the data acquisition and reduction,
while in section~\$\ref{sec:analysis} we report  the temporal and spectral analysis of the observed data. In section \$\ref{sec:model}, we present
the time-dependent model and compare its predictions with the results.
In section \$\ref{sec:disc} the results are summarised and discussed.

\section{Data Reduction}
\label{sec:data}

1ES~1959+650 was observed by \textit{AstroSat} during 25-26 October, 2017 for a total duration of $\sim$135~ks. \textit{AstroSat} \citep{2017JApA...38...27A}, being a multi-wavelength observatory equipped with Ultra-Violet Imaging Telescope (UVIT; \cite{2017JApA...38...28T, 2017AJ....154..128T}), Soft X-ray focusing Telescope (SXT; \cite{2017JApA...38...29S, 2016SPIE.9905E..1ES}), Large Area X-ray Proportional Counter (LAXPC; \cite{2016SPIE.9905E..1DY}) and Cadmium Zinc Telluride Imager (CZTI; \cite{2017arXiv171010773R}), allows the simultaneous monitoring of the source at UV, soft X-ray and hard X-ray bands. For this study we use the simultaneous observations of the source with LAXPC, SXT and UVIT. The observed and/or processed data for the observations were obtained from the \textit{AstroSat} data archive {\sc astrobrowse}. All the instrument specific software and tools for processing the data, and other important links are available in the ASSC website\footnote{http://astrosat-ssc.iucaa.in/?q=data{\_}and{\_}analysis}.

\begin{table}
\begin{center}
\tabcolsep 0.1cm
\caption{Details of the LAXPC, SXT and UVIT observations of 1ES\,1959+650 during October 2017 flaring.}
\begin{tabular}{cccc}
\hline
\hline
\\
\multicolumn{1}{c}{Instrument} & \multicolumn{1}{c}{Energy} & \multicolumn{1}{l}{Exposure} & \multicolumn{1}{c}{Count rate}\\
\multicolumn{1}{c}{} & \multicolumn{1}{c}{band} & \multicolumn{1}{c}{k\,sec} & \multicolumn{1}{c}{counts~s$^{-1}$} \\\hline
LAXPC10 & 3-30 keV & 60 & 60.80$\pm$0.21\\
LAXPC20 & 3-30 keV & 60 & 58.23$\pm$0.72\\
SXT & 0.7-7 keV & 30 & 13.98$\pm$0.02\\
UVIT & FUV-BaF2 (1541~\AA) & 14 & 4.37$\pm$0.03\\
 & NUV-N2 (2792~\AA) & 14 & 2.55$\pm$0.03\\
\hline
\hline
\end{tabular}
\label{tab_obs}
\end{center}
\end{table}

SXT is an X-ray imaging telescope that operates in the 0.3$-$8~keV energy band \citep{2017JApA...38...29S, 2016SPIE.9905E..1ES}. The source 1ES~1959+650 was observed in the Photon Counting (PC) mode with SXT. We processed the Level-1 data of the source through the SXT pipeline version 1.4a (AS1SXTLevel2-1.4a release date: December 06, 2017), using the calibration database (CALDB). The cleaned Level-2 event files for different orbits were merged with the {\sc sxtevtmerger} tool. We then used the merged event file to create the science products using {\sc xselect} version~V2.4d. SXT spectra and light curves were extracted for a circular region of the source with 16\,arcmin radius.

LAXPC consists of three units of identical X-ray proportional counters (LAXPC10, LAXPC20 \& LAXPC30) and can perform observation in a broad range of 3~keV to 80~keV \citep{2017ApJS..231...10A, 2017ApJ...835..195M, 2016ApJ...833...27Y, 2017JApA...38...30A}. LAXPC30 has been switched off due to gain instability issues \citep{2017ApJS..231...10A}, hence data from the other two LAXPC units were used for the analysis. The processing of Level-1 data and further analysis were done with the software {\sc laxpcsoft}. Since the source is faint above $\sim$30~keV in LAXPC we utilised the scheme of faint source background implemented as a part of {\sc laxpcsoft} for extracting the background spectra and light curves for LAXPC~10 \& LAXPC~20.

The source was observed simultaneously with UVIT. For UVIT analysis, we obtained the Level-2 data for the observation directly from {\sc astrobrowse}.  The Level-1 data were already processed with the pipeline {\sc  UVIT Level-2 Pipeline (UL2P)} version~6.3 (CALDB version - 20190625) by the payload operation team. The observation was made in PC mode in both FUV and NUV bands with BaF2 \& N2 filters, respectively. 

The details of the UV and X-ray observations of the source are given in Table~\ref{tab_obs}. Further analysis carried out for the study are given in the following sections.

\section{Analysis}
\label{sec:analysis}

\subsection{Time-Variability}
\label{sec:time}

\begin{figure*}
\begin{center}
\includegraphics[angle=0,scale=0.41]{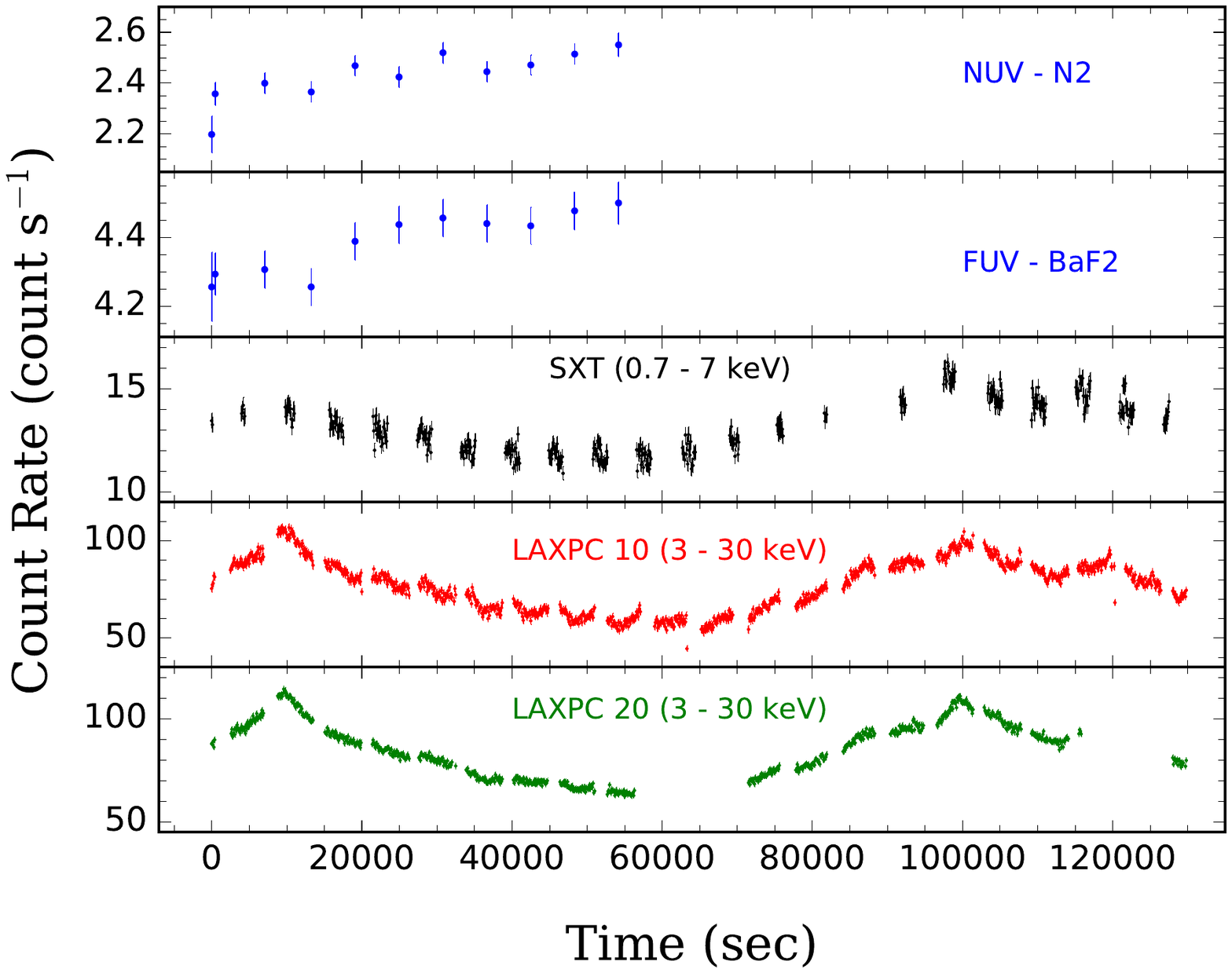} 
\includegraphics[angle=0,scale=0.41]{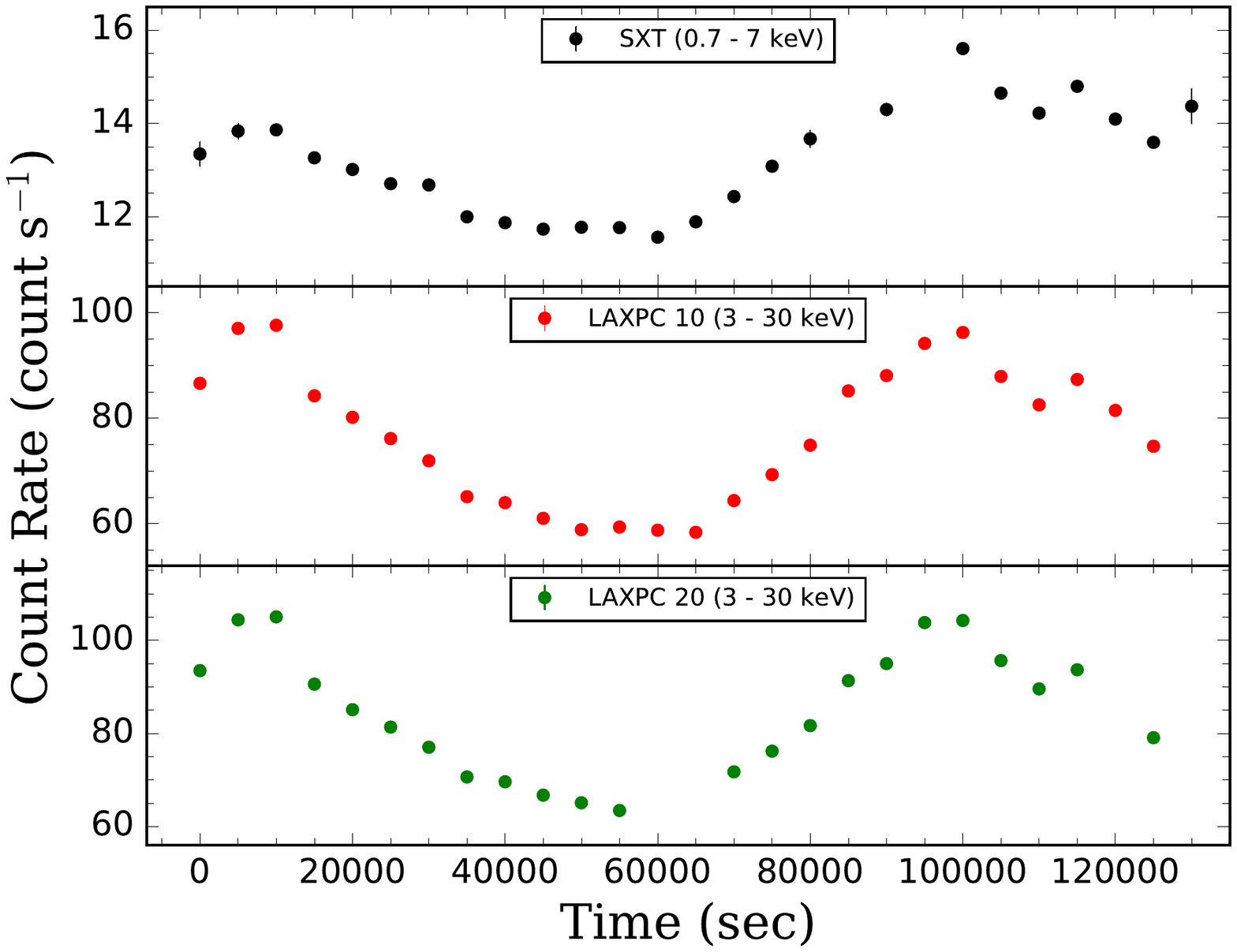} 
\vspace{-2.2cm}
\caption{ X-ray and UV light curves of 1ES\,1927+654 \textit{AstroSat} observations taken in October~2017. Left side: upper two panels show UVIT NUV-\textit{N2} and FUV-\textit{BaF2} observations, the middle panel shows the simultaneous 100 sec binned SXT light curve in the 0.7--7~keV band and the hard X-ray 100 sec binned light curves (3--30~keV) from LAXPC10 \& LAXPC20 observations are plotted in the lower panels. Right side: upper panel corresponds to the 5 ksec binned SXT light curve in the 0.7--7~keV band, middle and lower panel corresponds to the 5 ksec binned LAXPC10 \& LAXPC20 light curves (3--30~keV), respectively.}

\label{fig:lc}
\end{center}
\end{figure*} 

We generated the light curves from the SXT, LAXPC and UVIT observations of 1ES~1959+650 to check the time-variability of the source. Figure~\ref{fig:lc} shows the UV and X-ray light curves of 1ES\,1959+650. 

A total of 11 orbit data were available for both FUV and NUV bands.  We obtained the count rates from the combined Level-2 images of each orbit. The source count rates were extracted from circular regions of 30 pixels ($\sim 12.5''$) radius such that it includes $\sim 97\%$ of the source flux \citep{2017AJ....154..128T}. Background counts were extracted from multiple source free circles of 60-pixel radii, away from the source. The background subtracted FUV and NUV light curves plotted in Fig~\ref{fig:lc} show that 
the source is marginally variable in both NUV and FUV bands. In order to confirm if the source is variable we checked the light curves of two stars in the field. The non-variability of the stars show that the source is intrinsically variable.  

The SXT light curves were generated for different time bins in the 0.7$-$7\,keV band.  The quantum efficiency of SXT is poor at low energy \citep{2017JApA...38...29S}, hence we ignored the SXT spectra below 0.7 keV.    We used {\sc xselect} for creating the light curves, the middle panel in the left side of  Fig~\ref{fig:lc} shows the same for a bin size of 100\,sec, while upper panel in the right side of Fig~\ref{fig:lc} corresponds to 5 ksec binned SXT light curve. The variability of the light curves was checked with the {\sc ftool} {\sc lcstats} and we found that  the source is variable with a fractional rms variability, $F_{\rm rms}$ obtained in the 100\,sec and 5\,ksec binned light curves as 0.085$\pm$0.004 and 0.084$\pm$0.012, respectively. The LAXPC (10 \& 20) light curves were also extracted for a bin size of 100\,sec and 5\,ksec. In the LAXPC, the spectra above 30 keV was ignored due to strong background. The lower panels of  Fig~\ref{fig:lc} show these light curves in the 3--30\,keV energy band. We also calculated the $F_{\rm rms}$ of the source in LAXPC using {\sc lcsats}. The $F_{\rm rms}$  in the 100\,sec and 5\,ksec binned LAXPC\,10 light curves are obtained as 0.171$\pm$0.004 and 0.166$\pm$0.023, respectively. While in the 100\,sec and 5\,ksec binned light curves of LAXPC 20, the  $F_{\rm rms}$ values are obtained as 0.158$\pm$0.004 and 0.154$\pm$0.023, respectively. These results indicate that the source is variable in the LAXPC observations as well.

\begin{figure*}
\includegraphics[angle=0, scale=0.35]{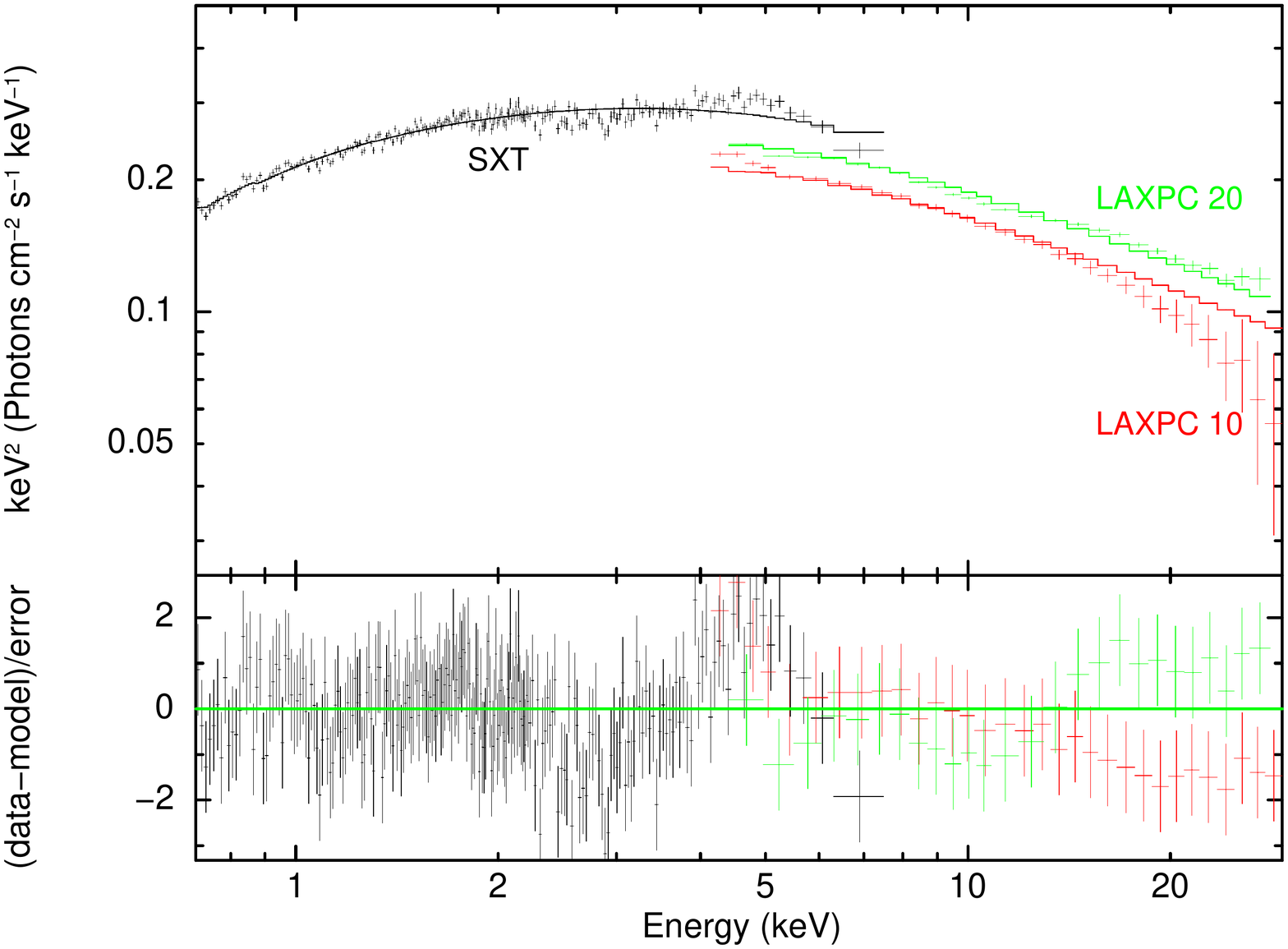}
\vspace{0.4cm}
\caption{The time-averaged SXT and LAXPC spectra (0.7--30~keV) of 1ES\,1959+650 fitted with the model \textit{constant$\times$TBabs}(\textit{Synconv$\otimes$bknpower}).
Upper panel: Unfolded spectra plotted with the model . Lower panel: Residuals for the spectral fit.}
\label{fig:spec_Tavg}
\end{figure*}


\begin{table}
\caption{ Best-fit parameters for the time-averaged spectra fitted with the synchrotron convolved \textit{power-law}, logparabola (\textit{logpar}) and broken-power-law (\textit{bknpower})  models. The top panel corresponds to synchrotron convolved \textit{power-law} model,  Row:- 5: power law particle index, p; and  6: normalisation, $n_{\rm pow}$.  Middle panel corresponds to synchrotron convolved \textit{logpar} model, Row:- 5: $\alpha$ is the particle spectral index at the pivot energy $\xi_{pivot}^2$; 6: $\beta$ is the curvature parameter;  7: $\xi_{pivot}$ determines the pivot energy; and 8: normalisation, $n_{\rm logpar}$. Bottom panel corresponds to synchrotron convolved \textit{bknpower} model,  Row:- 5: $\Gamma_1$ is the particle spectral index before the break energy; 6: $\Gamma_2$ is the particle spectral index after the break energy; 7: $\xi_{\rm brk}$ corresponds to the square root of break energy; and 8: normalisation,  n$_{\rm bkn}$. In each panel, Row 1 corresponds to hydrogen column density ($N_H$),  and Row 2, 3, 4  corresponds to the model constant values for  SXT, LAXPC 10, LAXPC 20 spectra, respectively;
}
\begin{tabular}{llll}
\hline
\hline
\multicolumn{1}{c}{\begin{tabular}[c]{@{}c@{}}Model\\ Component\end{tabular}} & \multicolumn{2}{c}{Parameters}                         & \multicolumn{1}{c}{$\chi^2$/dof} \\
\hline
\\
power-law                                                                      &  $N_{\rm H}$ ($10^{22}$cm$^{-2}$)                 & $\sim0.2$             & 1320.04/279                      \\
										& $\rm constant_{sxt}$		& 		1 	&		\\
										& $\rm constant_{lax10}$		&	$\sim 0.75$		&		\\		
										& $\rm constant_{lax20}$		&	$\sim 0.82$		&		\\
                                                                              & p                            & $\sim3.5$             &                                  \\
                                                                              & n$_{\rm pow}$                & $\sim2.5$             &                                  \\
\hline
\\
logpar                                                                        & $N_{\rm H}$ (10$^{22}$cm$^{-2}$)                 & 0.04$\pm$0.01          & 309.36/278                      \\
										& $\rm constant_{sxt}$		& 		1 	&		\\
										& $\rm constant_{lax10}$		&	$0.73_{-0.01}^{+0.01}$		&		\\		
										& $\rm constant_{lax20}$		&	$0.84_{-0.02}^{+0.02}$		&		\\

                                                                              & $\alpha$                     & 1.13$^{+0.16}_{-0.18}$          &                                  \\
                                                                              & $\beta$                      & 2.35$^{+0.18}_{-0.16}$          &                                  \\
                                                                              & $\xi_{pivot}$    ($\sqrt{keV}$)             & 0.7
                                                                              &                                  \\
                                                                              & n$_{\rm logpar}$ (10$^{-2}$) & 2.87$^{+0.22}_{-0.24}$          &                                  \\
                                                                            
\hline
\\
bknpower                                                                      & $N_{\rm H}$ (10$^{22}$cm$^{-2}$)                 & 0.04$\pm0.02$ & 286.73/277                       \\

										& $\rm constant_{sxt}$		& 		1 	&		\\
										& $\rm constant_{lax10}$		&	$0.74_{-0.02}^{+0.02}$		&		\\		
										& $\rm constant_{lax20}$		&	$0.85_{-0.02}^{+0.02}$		&		\\

                                                                              & $\Gamma_1$                           & 1.83$_{-0.36}^{+0.27}$ &                                  \\
                                                                              & $\Gamma_2$                           & 4.16$_{-0.09}^{+0.11}$ &                                  \\
                                                                              & $\xi_{\rm brk}$  ($\sqrt{keV}$)             & 1.69$_{-0.15}^{+0.14}$ 
                                                                              &                                  \\
                                                                              & n$_{\rm bkn}$                & 1.29$\pm0.03$
                                                                              &         \\   
\hline
\hline
\end{tabular}
\label{tab_Tavg}
\end{table}

\begin{figure*}
\begin{center}
\includegraphics[trim=0cm 0cm 0cm 0cm, clip=true, scale=0.49, angle=0]{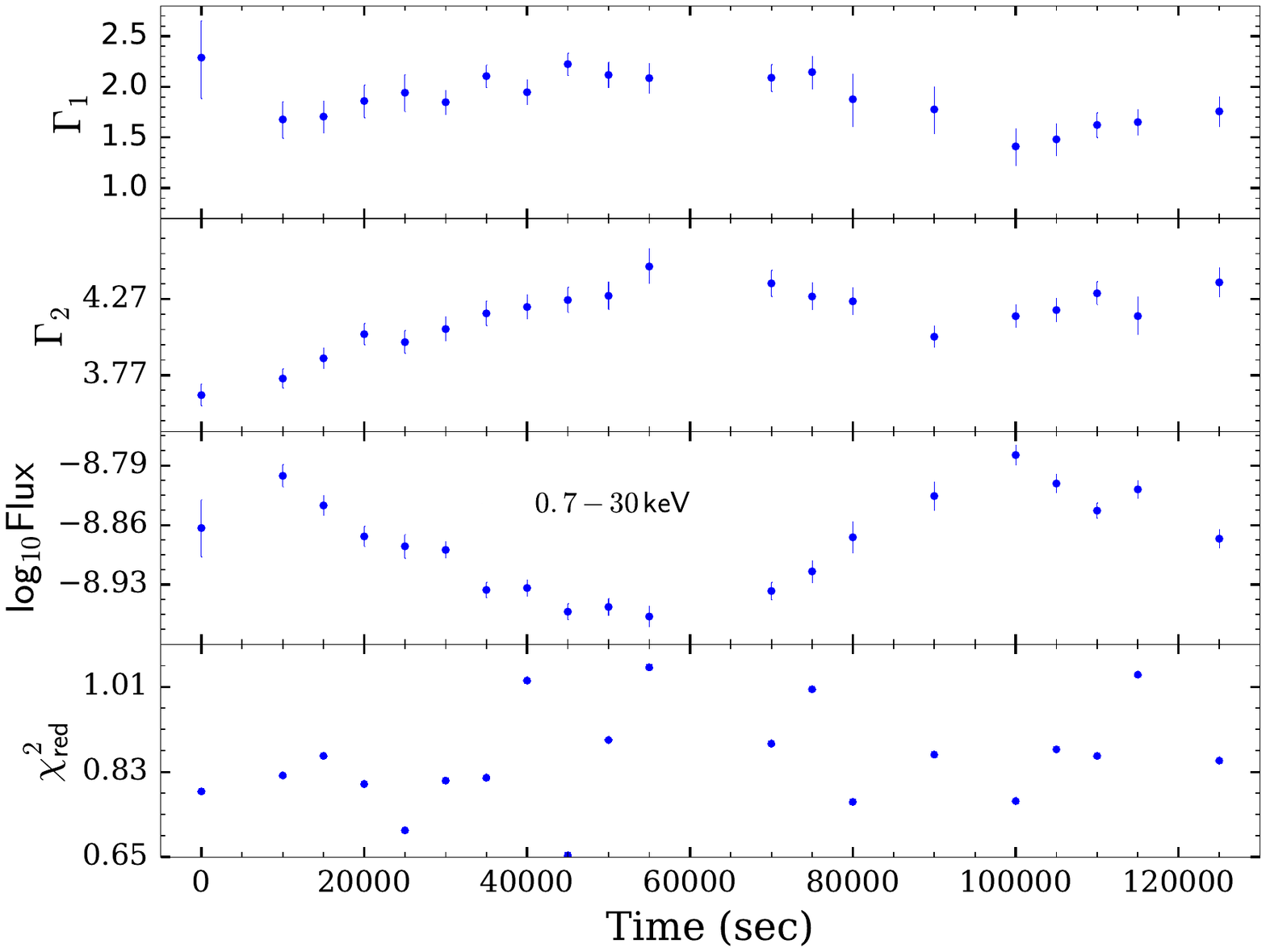}
\vspace{-2.5cm}
\caption{Best-fit results from the time-resolved spectral analysis using the model constant$\times$TBabs(Synconv$\otimes$bknpower). The first three panels show the variability of the particle indices ($\Gamma_1$ \& $\Gamma_2$) and the logarithm of flux ($\rm~erg~cm^{-2}~s^{-1}$) in the 0.7--30~keV band. The reduced $\chi^2$ value obtained for the spectral fit in each time-resolved segment is given in the lower panel.}
\label{Fig_par}
\end{center}
\end{figure*}

\begin{figure*}
\begin{center}
\includegraphics[trim=0cm 0cm 0cm 0cm, clip=true, scale=0.5, angle=0]{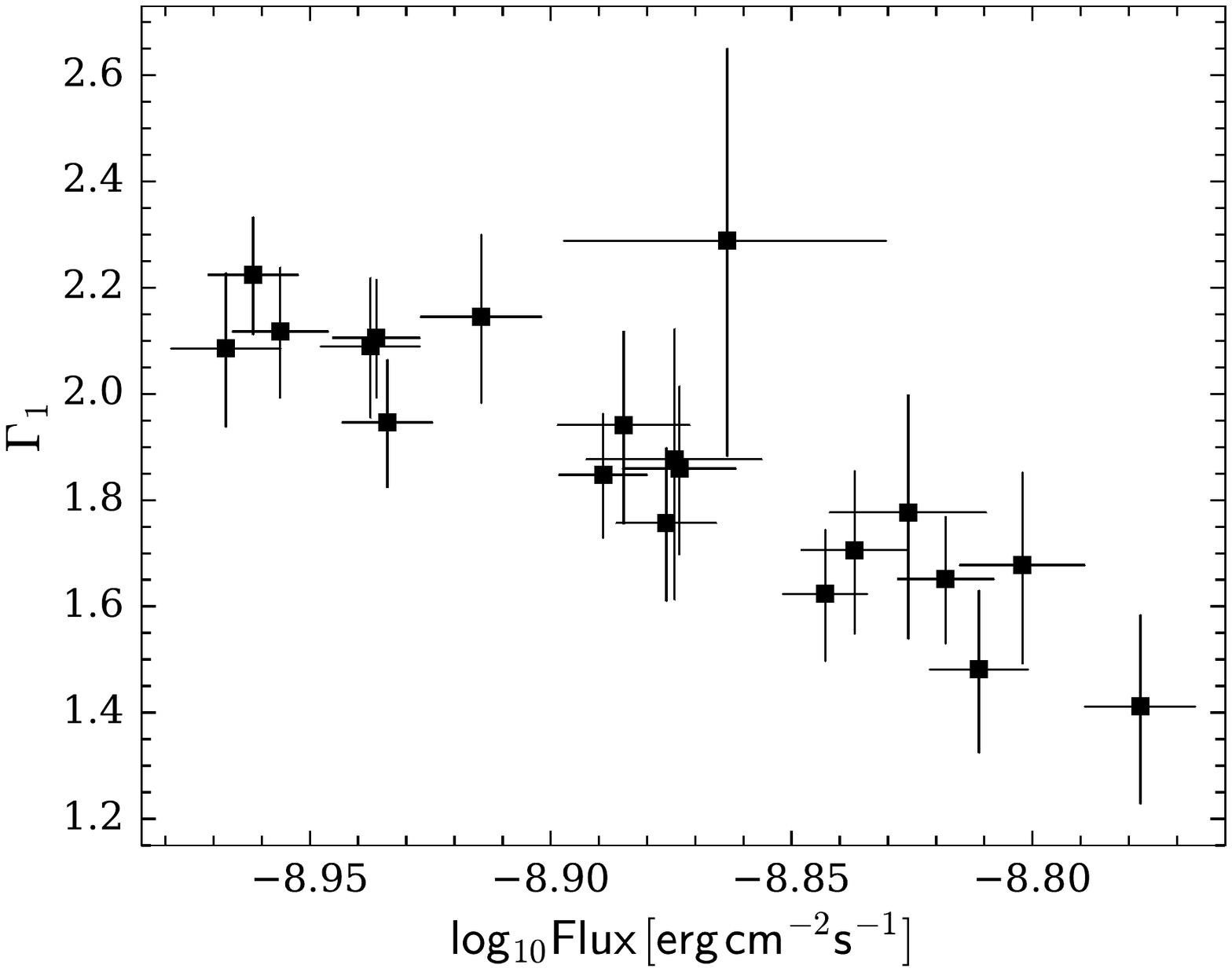}
\vspace{-2.5cm}
\caption{Variation of $\Gamma_1$ with Flux in the 0.7--30~keV band obtained from the time-resolved spectral analysis using the model \textit{constant$\times$TBabs}(\textit{Synconv$\otimes$bknpower}).}
\label{Fig_Fp1_corr}
\end{center}
\end{figure*}

\subsection{X-ray spectral analysis}
\label{sec:spec}

We analysed the X-ray spectra of the source using XSPEC version~12.9.0n. The X-ray spectra were analysed in the energy band of $\sim$ 0.7$-$30\,keV, where the SXT spectral data were included from 0.7\,keV to $\sim$7\,keV and LAXPC data were ignored below 4~keV and above $\sim$ 30\,keV. The background spectrum for SXT was produced from blank-sky observations by SXT POC. For LAXPC, the background spectra were generated using the code for the faint source background provided by LAXPC team. We have used sxt\_pc\_mat\_g0to12.rmf as SXT response matrix file (RMF). The ancillary response file (ARF) for the SXT instrument was obtained using the standard PC mode ARF and the sxt\_ARFModule Version: 0.02 released in 2019 July.

Throughout the fitting procedure we applied a systematic error of 3\%. While fitting the SXT spectrum we need to incorporate the energy shift in the response matrix. For this we used {\emph gain fit} command with the slope frozen at 1, and the offset was obtained around 0.03~keV. The SXT spectrum was grouped such that each bin has minimum 1000~counts and LAXPC spectra were grouped at 5\% level to obtain three energy bins per resolution. The time-averaged spectra of SXT, LAXPC~10 and LAXPC~20 were fitted simultaneously with three different models, power-law, log-parabola and broken power-law, corrected for the Galactic absorption using the XSPEC routine \textit{TBabs}.
A constant was multiplied to the SXT spectra  to take into account any systematic variation in the effective areas of the SXT and LAXPC units.
These models were convolved with the synchrotron emissivity function \citep{rybicki2008radiative} using an XSPEC local convolution model \textit{Synconv} written by us. Once convolved with \textit{Synconv} these models represent particle distribution of the emitting region instead of photon distribution. Specifically the ``Energy'' variable in the convolved XSPEC model should be interpreted as being $\xi = \sqrt{C} \gamma$, where $\gamma$ is the Lorentz factor of the particle and $C = 1.36\times 10^{-11} \delta B/(1+z)$ with $\delta$ being the Doppler factor, $z$ the red-shift and $B$ the magnetic field strength. Thus, any parameter of the model, such as the break energy in the broken power-law model $\xi_{brk}$, would
be related to $\gamma$ as $\gamma_{brk} = \xi_{brk}/\sqrt(C)$ and one can relate a corresponding break in the photon energy spectrum $E_{brk} = C \gamma_{brk}^2 = \xi_{brk}^2$.
The spectral fits with the synchrotron convolved power-law, log-parabola
and broken power-law models resulted in $\chi^2$(/dof) values of 1320.04(/279), 309.36(/278) and 286.73(/277) 
respectively. The best-fit parameters of these models fitted for the time-averaged spectra are given in Table~\ref{tab_Tavg}.  Due to the very large $\chi^2$ we discarded \textit{power-law} in the later analysis. The broken power-law provides a better fit compared to the other models and the spectral fit and residual for the model \textit{constant$\times$TBabs(Synconv$\otimes$bknpower)} is shown in Fig.~\ref{fig:spec_Tavg}. The flux obtained for this model in the 0.7--30~keV range is $\sim 1.4\times10^{-9}{\rm erg~cm^{-2}s^{-1}}$.  The spectral fit obtained by freezing the hydrogen column density, $N_H$  to the value given in LAB survey \citep{2005A&A...440..775K} resulted in higher $\chi^2$ values than those acquired when $N_H$ was kept free. 
Therefore, in order to obtain better fit statistics, the $N_H$ was set as a free parameter. The obtained values  of $N_H$ are lower than those obtained from LAB survey group.

Since the source is highly variable in the soft and hard X-ray bands we investigated the time evolution of X-ray spectral properties of the source during observation. Here, we employed the method of time-resolved spectroscopy, where the total observation period was divided into time segments of 5\,ksec. For each of the time segment we created SXT and LAXPC spectra. In this analysis, we used only SXT and LAXPC~20 spectra because LAXPC~20 background is more stable than LAXPC~10.

Since the broken power-law yielded a better fit statistic than the log-parabola one for the time averaged spectrum, we use it for the time resolved analysis. However, the break energy $E_{\rm brk}$ (or $\xi_{brk}^2$) , was not well constrained and hence we fix its value to $ 2.86$\,keV obtained during the time-averaged spectral fitting.  The $\xi_{\rm brk}$ and $N_H$ parameters were fixed at the values obtained for the time-average spectral fit where the relative error in normalisation is minimum. We note that since the model is being convolved with synchrotron emissivity function, the indices of the broken power-law model represent the indices of the particle distributions i.e. $\Gamma_1$ and $\Gamma_2$ are the indices of power law distributions before and after the break energy.

The best-fit parameters for the time-resolved spectral analysis are plotted in Figure.~\ref{Fig_par} and are also given in Table \ref{table:trs}. A strong anti-correlation (Spearman's method; rank $\sim$ -0.8, p-value $\sim 2.5\times10^{-5}$) is observed between the flux and $\Gamma_1$ (see Fig.~\ref{Fig_Fp1_corr}), which is the hardening when brightening behaviour previously reported in blazars. 

\begin{table*}
\caption{Best fit parameters of time resolved spectral analysis fitted with the synchrotron convolved broken-power-law models.  Col. 1: time bin in secs; 2: low energy particle spectral index; 3: high energy particle spectral index;  4: normalisation;  5: logarithm of Flux ($\rm~erg~cm^{-2}~s^{-1}$); 6: $\chi^2$ / degrees of freedom}
\begin{adjustbox}{width=0.7\textwidth,center=\textwidth}
\begin{tabular}{lcccccr}
\hline
Time  &  $\Gamma_1$  &  $\Gamma_2$  &  $n_{bkn}$  &  logF   &  $\chi^2 / dof $\\\\
\hline
0 & 2.29$^{+0.36}_{-0.41}$ & 3.64$^{+0.07}_{-0.07}$ & 1.19$^{+0.06}_{-0.07}$ & -8.86$^{+0.03}_{-0.03}$  & 52.87 / 67 \\\\
10000 & 1.68$^{+0.18}_{-0.19}$ & 3.75$^{+0.06}_{-0.06}$ & 1.19$^{+0.04}_{-0.04}$ & -8.8$^{+0.01}_{-0.01}$  & 101.24 / 123 \\\\
15000 & 1.71$^{+0.15}_{-0.16}$ & 3.88$^{+0.06}_{-0.06}$ & 1.16$^{+0.03}_{-0.03}$ & -8.84$^{+0.01}_{-0.01}$  & 134.85 / 156 \\\\
20000 & 1.86$^{+0.15}_{-0.16}$ & 4.04$^{+0.07}_{-0.07}$ & 1.18$^{+0.03}_{-0.03}$ & -8.87$^{+0.01}_{-0.01}$  & 106.22 / 132 \\\\
25000 & 1.94$^{+0.18}_{-0.19}$ & 3.99$^{+0.08}_{-0.07}$ & 1.15$^{+0.03}_{-0.03}$ & -8.88$^{+0.01}_{-0.01}$  & 81.29 / 115 \\\\
30000 & 1.85$^{+0.12}_{-0.12}$ & 4.07$^{+0.08}_{-0.08}$ & 1.14$^{+0.02}_{-0.02}$ & -8.89$^{+0.01}_{-0.01}$ & 166.48 / 205 \\\\
35000 & 2.11$^{+0.11}_{-0.11}$ & 4.18$^{+0.08}_{-0.08}$ & 1.13$^{+0.02}_{-0.02}$ & -8.94$^{+0.01}_{-0.01}$ & 105.54 / 129 \\\\
40000 & 1.95$^{+0.12}_{-0.12}$ & 4.22$^{+0.08}_{-0.08}$ & 1.1$^{+0.02}_{-0.02}$ & -8.93$^{+0.01}_{-0.01}$  & 119.79 / 117 \\\\
45000 & 2.22$^{+0.11}_{-0.11}$ & 4.26$^{+0.08}_{-0.08}$ & 1.13$^{+0.02}_{-0.02}$ & -8.96$^{+0.01}_{-0.01}$ & 75.85 / 116 \\\\
50000 & 2.12$^{+0.12}_{-0.13}$ & 4.29$^{+0.09}_{-0.09}$ & 1.12$^{+0.02}_{-0.02}$ & -8.96$^{+0.01}_{-0.01}$  & 105.07 / 117 \\\\
55000 & 2.09$^{+0.14}_{-0.15}$ & 4.49$^{+0.12}_{-0.11}$ & 1.13$^{+0.02}_{-0.02}$ & -8.97$^{+0.01}_{-0.01}$ & 105.17 / 100 \\\\
70000 & 2.09$^{+0.13}_{-0.13}$ & 4.37$^{+0.09}_{-0.09}$ & 1.18$^{+0.02}_{-0.02}$ & -8.94$^{+0.01}_{-0.01}$  & 144.21 / 162 \\\\
75000 & 2.15$^{+0.16}_{-0.16}$ & 4.29$^{+0.09}_{-0.09}$ & 1.24$^{+0.03}_{-0.03}$ & -8.91$^{+0.01}_{-0.01}$  & 119.63 / 119 \\\\
80000 & 1.88$^{+0.25}_{-0.27}$ & 4.26$^{+0.09}_{-0.09}$ & 1.26$^{+0.05}_{-0.05}$ & -8.87$^{+0.02}_{-0.02}$  & 63.66 / 83 \\\\
90000 & 1.78$^{+0.22}_{-0.24}$ & 4.02$^{+0.07}_{-0.07}$ & 1.27$^{+0.05}_{-0.05}$ & -8.83$^{+0.02}_{-0.02}$ & 79.77 / 92 \\\\
100000 & 1.41$^{+0.17}_{-0.18}$ & 4.16$^{+0.07}_{-0.07}$ & 1.33$^{+0.04}_{-0.05}$ & -8.78$^{+0.01}_{-0.01}$  & 109.94 / 143 \\\\
105000 & 1.48$^{+0.15}_{-0.16}$ & 4.2$^{+0.08}_{-0.07}$ & 1.27$^{+0.03}_{-0.04}$ & -8.81$^{+0.01}_{-0.01}$  & 156.31 / 178 \\\\
110000 & 1.62$^{+0.12}_{-0.13}$ & 4.31$^{+0.08}_{-0.08}$ & 1.27$^{+0.03}_{-0.03}$ & -8.84$^{+0.01}_{-0.01}$  & 181.49 / 210 \\\\
115000 & 1.65$^{+0.12}_{-0.12}$ & 4.16$^{+0.13}_{-0.12}$ & 1.3$^{+0.02}_{-0.03}$ & -8.82$^{+0.01}_{-0.01}$  & 232.13 / 224 \\\\
125000 & 1.76$^{+0.14}_{-0.15}$ & 4.38$^{+0.1}_{-0.09}$ & 1.25$^{+0.03}_{-0.03}$ & -8.88$^{+0.01}_{-0.01}$  & 146.95 / 172 \\\\
\bottomrule
\end{tabular}
\end{adjustbox}
\label{table:trs}
\end{table*}

 During Oct 25-26, 2017 \emph{Swift} has carried one observation for an exposure of 987 secs in two orbits in the Windowed timing (WT) mode. However, during the first orbit, the image is too near the edge of the WT window, which resulted in loss of photon counts, hence we skipped the first orbit in our analysis. We obtained the spectrum for the second orbit having an exposure of 499 sec using an online tool available at the UK \emph{Swift} Science Data Centre \citep{2009MNRAS.397.1177E}. The second orbit falls in the 15th time-segment (corresponding time 70 ksec) of our time resolved spectral analysis. We carried out the simultaneous spectral fit of the \emph{Swift}-XRT (0.7--7 keV), SXT (0.7--7 keV) and LAXPC (3--30 keV) spectrums with the synchrotron convolved broken power law model. The model fitted the spectrum well with a reduced-$\chi^2$ of 1.01, and the best fit parameters are obtained as $\Gamma_1=2.19_{-0.24}^{+0.22}$, $\Gamma_2=4.36_{-0.09}^{+0.09}$ and $n_{bkn}=1.04_{-0.02}^{+0.02}$. The obtained spectral index, $\Gamma_1$  is consistent with that acquired in the time resolved spectral analysis (see Table \ref{table:trs}).

\subsection{Correlation between particle density and Index }
\label{sec:corr}

\begin{figure*}
\begin{center}
\includegraphics[scale=0.64, angle=0]{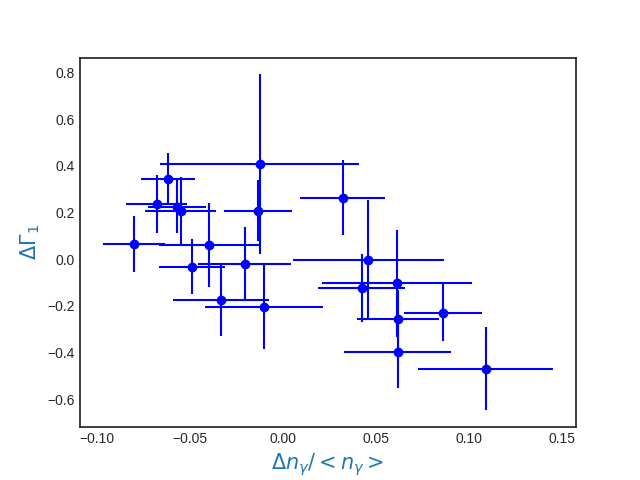}
\caption{Plot between the variation of $\Delta \Gamma_1$ with the normalised particle density variation $\Delta n_\gamma(\xi_{ref})/<n_\gamma(\xi_{ref})>$ of the model \textit{constant$\times$TBabs}(\textit{Synconv$\otimes$bknpower}) fitted to the time-resolved spectral bins.}
\label{Fig_Np1_corr}
\end{center}
\end{figure*}

		\begin{figure*}
		\begin{center}
        \includegraphics[scale=0.64]{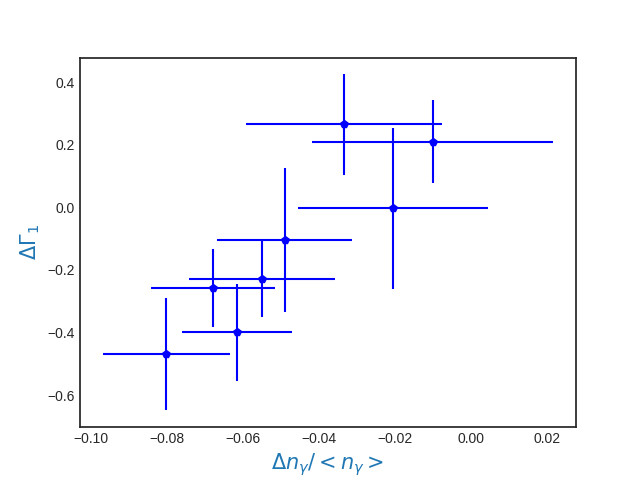}

        \caption{Plot between  $\Delta\Gamma_1$ and $\Delta n_\gamma(\xi_{ref})/<n_\gamma(\xi_{ref})>$ when the $\Gamma_1$ is shifted by 60~ksec.
}       
        \label{fig:pos_corel}
		\end{center}        
        \end{figure*}

The X-ray variability in flux and spectral parameters of 1ES\,1959+650 observed by \textit{AstroSat} can be used to obtain information about the physical parameters of the emission region. Theoretically, it is easier to obtain the predicted  relation between index and the particle density at a given energy rather than between the index and  the flux. Thus we also estimated the particle density variation.

\begin{figure*}
		\begin{center}
        \includegraphics[scale=0.7]{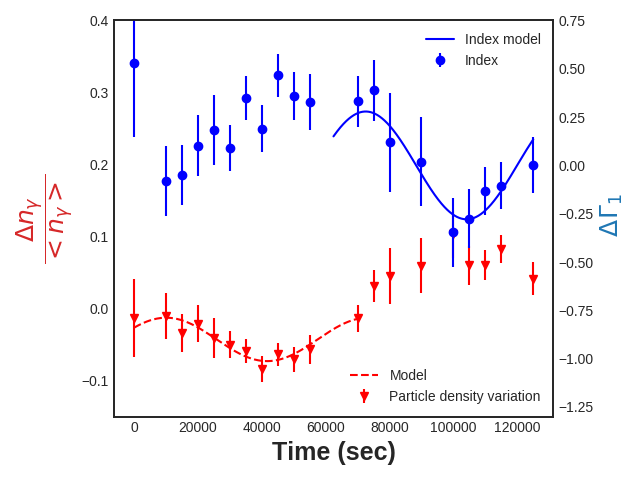}
        \caption{ $\Delta n_\gamma/n_\gamma$ and $\Delta\Gamma_1$  light curves showing a time delay of nearly $\sim$60 ks.  The red dotted curve corresponds to input sinusoidal variation of $\frac{\Delta n_c}{n_c}$ (equation \ref{eq:sino_var}), while the blue solid curve is the corresponding index variation obtained using equation \ref{eq:time_lag}.}       
        \label{fig:time_delay_bb}
		\end{center}        
        \end{figure*}

The normalisation of the broken power-law model is such that it is equal to the particle density at the break energy, i.e. at $\xi_{brk} = \sqrt{C} \gamma_{brk}$, such that $E_{brk} = \xi_{brk}^2$. However this is a matter of choice and the model can be recasted such that the normalisation is equal to the density at any chosen particle energy $\xi$. For the time-averaged spectrum we computed the normalisation and its error for different values of $\xi$ and  found that for $\xi^2~(or~ E)= 1.21 \,keV$, the error on the normalisation i.e. the particle density at $\xi^2 = 1.21\, keV$, is the smallest. Hence we took this value as the reference energy. Note that this energy is smaller than the break one and hence we look for correlation between the normalised particle density variation
 $\Delta n_\gamma(\xi_{ref})/<n_\gamma(\xi_{ref})>$ (where $<n_\gamma(\xi_{ref})>\,= 1.20$)   and the  variation in lower energy index i.e. $\Delta \Gamma_1$  (i.e. deviation of $\Gamma_1$ from their means $1.88$) as shown in 
Figure \ref{Fig_Np1_corr}. An anti-correlation is seen (Spearman rank correlation $\sim -0.68$  with probability $0.001$) which
is not as strong as that between the flux and the index (Spearman rank correlation $\sim -0.8$ with probability $2.5\times 10^{-5}$). This is perhaps expected since the definition of flux makes it a function of the normalisation of the particle density  and index, and some of the observed anti-correlation can be attributed to this dependence on index. It is interesting to note that a better correlation is obtained when the index is shifted by $ 60 $ ksecs (Figure \ref{fig:pos_corel}), i.e. a time-lag is introduced between  them  (Spearman rank correlation $\sim 0.9$  with probability $0.002$).  In other words, a strong correlation is obtained between the low-energy index and particle density for segments which are 60 ksec apart. This can be  qualitatively illustrated by plotting the normalised particle density variation  $\Delta n_\gamma(\xi_{ref})/<n_\gamma(\xi_{ref})>$ and $\Delta \Gamma_1$  as a function of time as shown in Figure \ref{fig:time_delay_bb}. While we discuss the implication of this behaviour in the next section, at this stage it is worth noting that the amplitude of the variation of particle density  $|\Delta n_\gamma(\xi_{ref})/<n_\gamma(\xi_{ref})| \sim 0.1$ is significantly smaller than for the index $|\Delta \Gamma_1| \sim 0.5$.

\section{Modeling the Spectral variability}
\label{sec:model}

The X-ray emission of HBL\, sources  is typically modelled as synchrotron emission, which is produced by a non-thermal electron distribution gyrating in an ambient magnetic field. The emission is most likely to be associated with a shock, which provides favourable conditions for the particles to get accelerated through the Fermi acceleration process. Keeping these features in mind, we investigated the predictions of such a model for the variability on the index  $\Gamma_1$ and its relation with that of the particle number density.

We assume that the observed emission arises from a relativistic spherical blob which is moving towards the line of sight of the observer with Doppler factor $\delta=\frac{1}{\Gamma(1-\beta\cos\theta)}$, where $\Gamma$ is the bulk Lorentz factor of the blob and $\theta$ is the angle between the jet axis and line of sight of observer. The spherical blob includes an  
acceleration region( which includes a shock) where injected low energy electrons are continuously accelerated to high energies.  The electrons escaping from the acceleration region enter into a cooling region, where they  cool through synchrotron and Inverse Compton processes.  We assume that the acceleration and the cooling region are spatially separated and that the escape rate from the acceleration region is  equal to the injection rate in the cooling region. For such a model, the steady state particle distribution in the cooling region is a broken power-law as assumed in the spectral fitting in the previous section. For particles with energy greater than the break energy, radiative cooling is important while for lower energy ones cooling is insignificant. Since the motivation here is to predict the correlation between the normalisation and index below the break energy, we consider the case when radiative cooling is not important.

 In the acceleration region, the time evolution of the particle distribution can be written in-terms of a partial differential  equation as 

\begin{equation}\label{eq:KE_acc}
\frac{\partial n_a}{\partial t}+ \frac{\partial}{\partial \gamma}\left\{\left[\left(\frac{d\gamma} {dt}\right)_{acc}-\left(\frac{d\gamma} {dt}\right)_{loss}\right]n_a\right\}+\frac{n_a}{\tau_{esc}}=0 
\quad ,
\end{equation}
where $\gamma$ is Lorentz factor of electron,  $n_a$ is  number density (in units of $cm^{-3}$) of the electrons in the acceleration region, first and second term in square bracket accounts for the acceleration rate and energy loss rates respectively, and $\frac{n_a}{\tau_{esc}}$ describes the escape from the acceleration region at rate $\tau_{esc}^{-1}$. 
 The acceleration term in the equation is approximated by

\begin{equation}\label{eq:acc_term}
\left(\frac{d\gamma}{dt}\right)_{acc}=\frac{\gamma}{\tau_{acc}}
\end{equation}
For synchrotron cooling, the energy-loss rate is given by
\begin{equation}\label{eq:loss_term}
\left(\frac{d\gamma}{dt}\right)_{loss}=\beta_s\gamma^2
\end{equation}
where
$\beta_s\sim\frac{1}{\gamma_{max}\tau_{acc}}=\frac{4}{3}\frac{\sigma_TB^2}{8\pi m_ec}$. Here  $B$ is the magnetic field in the  acceleration region, $\sigma_T$ is the  Thomson cross section and  $m_e$ is the electron mass.

The accelerated particles finally escape into the cooling region, where they subsequently lose most of energy into radiation. Since, as mentioned above we are interested in the spectra below the break energy, we ignore the cooling term in this region and hence we solve the kinetic equation of particle distribution for $\gamma\lesssim {\gamma_{brk}}$. In such cases, the  evolution of particle distribution in the cooling region can be described by the equation 
\begin{equation}\label{eq:KE_CR}
	\frac{\partial n_c}{\partial t}=\frac{n_c}{\tau_{esc,c}}+\frac{n_a}{\tau_{esc}}
\end{equation}
Note that in this interpretation the observed synchrotron spectrum is produced by the non-thermal electrons in the
cooling region given by $n_c(\gamma)$.

The steady state solution of equation \ref{eq:KE_acc} can be obtained as 
\begin{equation}\label{eq:ks}
\begin{split}
n_a(\gamma)&=K\,\gamma
_0^{\frac{\tau_{acc}}{\tau_{esc}}}\left(1-\frac{\gamma_0}{\gamma_{max}}\right)^{-\frac{\tau_{acc}}{\tau_{esc}}}\\
&\times \frac{1}{\gamma^2}\left(\frac{1}{\gamma}-\frac{1}{\gamma_{max}}\right)^{\frac{\tau_{acc}}{\tau_{esc}}-1}
\end{split}
\end{equation}
 where $K$ is constant and can be determined by boundary condition at some $\gamma_0$ or by introducing the  mono-energetic injection $Q=Q_0\delta(\gamma-\gamma_0)$ in equation \ref{eq:KE_acc}. In later case (mono-energetic injection), $K$ is obtained as $Q_0 \tau_{acc}$. 
 Here we  consider the case when the solution of equation \ref{eq:KE_acc} is determined
 by the boundary condition, while the alternate situation of a delta function injection is discussed in the end of this section. For  $\gamma\ll\gamma_{max}$ and  $\gamma_0\ll\gamma_{max}$, the equation reduces to a power-law form,
\begin{equation}
  n_a(\gamma)=\frac{K}{\gamma_0}\,\left(\frac{\gamma}{\gamma_0}\right)^{-\Gamma}
\end{equation}
with $\Gamma=\frac{\tau_{acc}}{\tau_{esc}}+1$ being the index. The corresponding steady state solution for the particle
density in the cooling region becomes
\begin{equation}
  n_c(\gamma)=\frac{K}{\gamma_0}\, \frac{\tau_{esc,c}}{\tau_{esc}}\,\left(\frac{\gamma}{\gamma_0}\right)^{-\Gamma}
\end{equation}

Now a variation in the acceleration time-scale $\Delta \tau_{acc}$ would lead to a variation in the
index $\Delta \Gamma = \Delta \tau_{acc}/\tau_{esc}$. If the variation is a slow process such that it is signficantly longer than
the time-scales of the system, then the particle density would vary as
\begin{equation}
  \frac{\Delta n_c(\gamma)}{n_c (\gamma)} = - \log \left(\frac{\gamma}{\gamma_0}\right) \, \Delta \Gamma
  \label{delta_n_steady}
\end{equation}
This implies that the density at some energy $\gamma$ will be inversely correlated with the index as is observed.
However, since $\gamma$ is expected to be significantly larger than $\gamma_0$, the amplitude of the density
variation $|\Delta n_c(\gamma)/n_c (\gamma)| \gg |\Delta \Gamma|$, which is contrary to observations. Figure \ref{fig:time_delay_bb} shows that
while  $|\Delta n_c(\gamma)/n_c (\gamma)| \sim 0.1$, $\Gamma$ shows a much larger variation with  $\Delta \Gamma \sim 0.5$. 
Another way of expressing the problem, is that the power-law  particle distribution pivots over $\gamma_0$ and since
$\gamma \gg \gamma_0$, the density at $\gamma$ should vary significantly for even small changes in index. However,
the data shows that the magnitude of the variation of the density is smaller than that of the index. We note that
the problem persists even when we consider the driving variability to be due to the escape time scale $\Delta \tau_{esc}$
or if we consider  mono-energetic injection such that $K = Q_0 \tau_{acc}$. 

A possible solution maybe that the observed variability is not a slow variation and instead its time-scale is comparable to the time-scales of the systems such as the acceleration time-scale. In such cases, the steady state solutions are no longer valid and one has to solve the time dependent equations \ref{eq:KE_acc} and \ref{eq:KE_CR}. We do this in the linear
approximation and introduce a small  sinusoidal perturbation in particle acceleration time scale 
\begin{equation}\label{eq:pert_esc}
\tau_{acc}(t)=\tau_{acc}+\Delta \tau_{acc}\,e^{\iota \omega t}
\end{equation}
which in the linear regime produce variation in the particle densities, $n_a(\gamma,t)=n_a(\gamma)+\Delta n_a e^{\iota \omega t}$ and $n_c(\gamma,t)=n_c(\gamma)+\Delta n_c e^{\iota \omega t}$. In such a case, the fractional change in particle density in the acceleration region for the sinusoidal perturbation in $\tau_{acc}$  is obtained as
\begin{equation}\label{eq:density_acc}
	\frac{\Delta n_a}{n_a(\gamma)}=\frac{\Delta \tau_{acc}}{\iota \omega \tau_{acc}\tau_{esc}}\left[\left(\frac{\gamma_0}{\gamma}\right)^{\iota\omega\tau_{acc}}-1\right]
\end{equation}
and the corresponding variation in the cooling region for energies lower than the break energy, becomes
\begin{equation}\label{eq:density_cool}
	\frac{\Delta n_c}{n_c(\gamma)}=\frac{\Delta \tau_{acc}}{\iota\omega(1-\iota \omega \tau_{esc,c})(\tau_{acc}\tau_{esc})}\left[\left(\frac{\gamma_0}{\gamma}\right)^{\iota\omega\tau_{acc}}-1\right]
\end{equation}
Now one can define a local particle density index as $\Gamma = -\left(\frac{\gamma}{n_c}\right)\frac{d n_c}{d\gamma}$ and its
variation, $\Delta \Gamma$, can be written as
\begin{equation}\label{eq:index_norm_cool}
\begin{split}
\Delta \Gamma&=-\gamma\frac{\partial}{\partial \gamma} \left(\frac{\Delta n_c}{n_c(\gamma)}\right)\\
&=\frac{\omega \tau_{acc}}{2}\frac{1}{\sin\left(\omega \tau_{acc}\log(\gamma_0/\gamma)/2\right)}\\
&\times\exp(\iota\omega \tau_{acc}\log(\gamma_0./\gamma)/2) \frac{\Delta n_c}{n_c(\gamma)}
\end{split}
\end{equation}

At this stage, it is convenient to introduce a real sine function perturbation, instead of complex quantities
and express the variation 
\begin{equation}{\label{eq:sino_var}}
\frac{\Delta n_c}{n_c}=N\sin(\omega t-\phi)
\end{equation}
where N is some  constant. The corresponding  index variation can then be written as
\begin{equation}\label{eq:time_lag}
\Delta\Gamma=-\frac{N}{\log(\gamma/\gamma_0)}\frac{\phi_{lag}}{\sin(\phi_{lag})}\sin(\omega t-\phi-\phi_{lag}) 
\end{equation}
where $\phi_{lag} = \omega \tau_{acc}\log(\gamma/\gamma_0)/2$ is the phase lag corresponding to a time
delay  $\tau_{lag}=\tau_{acc}\log(\gamma/\gamma_0)/2$ between the index and normalisation. It is evident that for slow variations, with $\omega$ tending to zero, $\phi_{lag}/\sin(\phi_{lag})$ tends to unity, and the above equation reduces
to equation  \label{delta_n_steady} derived from the steady state solution. Now, when the variation time-scale is
comparable to the acceleration one, i.e. when $\phi_{lag}$ is not small, the term  $\phi_{lag}/\sin(\phi_{lag})$ can be
arbitrarily large. This means that the amplitude of $\Delta\Gamma$ can be much larger than that of $\frac{\Delta n_c}{n_c}$. However, in that case there will be a significant time-lag between the variation of the normalisation and the index. The equations have been derived in the rest frame of the emitting region. In the observer's frame the change in the frequency would be $\omega_{o} = \omega \delta/(1+z)$, where $z$ is the redshift and $\delta$ is Doppler factor.

We now check, if the observed variation of particle density  and the index can be understood in this framework.
The values of N, $\omega_o$ and $\phi$ are chosen such that equation \ref{eq:sino_var} produces the shape of the particle density light curve as shown in Figure (\ref{fig:time_delay_bb}), where the red dashed curve is the input sinusoidal variation (equation \ref{eq:sino_var}) for $N=3\times10^{-2}$, $\omega_o=9.94\times10^{-5}$ and $\phi=5.71$.  The blue solid curve is the corresponding index variation obtained by using equation (\ref{eq:time_lag}) with
$\phi_{lag} = 6.21$  and $\log{\gamma/\gamma_0}=9.05$ after the expected time-delay of  $\tau_{lag,o} = \phi_{lag}/\omega_o \sim 6.24\times10^4$ seconds. This further validates the correlation seen between the density and index after taking into account a time-lag shift as shown in Figure \ref{fig:pos_corel}. Note that in the emission rest frame,  $\tau_{lag} = \phi_{lag}/\omega \sim 6.24\times10^4\delta/(1+z)$ secs and the corresponding $\tau_{acc} = 1.38\times10^4 \delta/(1+z) $ secs.

On can undertake a similar analysis for the case of when instead of having a fixed low energy boundary condition for the particle density, there is a mono-energtic injection of particles i.e. $Q=Q_0\delta(\gamma-\gamma_0)$ in equation \ref{eq:KE_acc}, leading to  $C = Q_0 \tau_{acc}$.  The analysis can also be done for the situation where the variability is driven by variation in the escape time-scale instead of the acceleration one. For both these cases, the variation of index $\Delta \Gamma$ turns out to be always significantly smaller than that of the density for any oscillation frequency. Thus, the observed large variation of $\Gamma$ as compared to the density is compatible only with a model with acceleration time scale variability and a fixed low energy boundary condition for the particles.

We note and emphasis again in the next section, that the variability observed is only for a single cycle and hence the time delay inferred between the density and index can not be asserted with statistical significance. The point here is that the observed magnitudes of the density and index can be explained if the time-scale of variation is of the same order as the acceleration time-scale and the expected time-lag between them is consistent with the data.

\section{Summary \& Discussion}
\label{sec:disc}

We have analysed the X-ray variability of 1ES\,1959+650 observed  during 26 October, 2017 by \textit{AstroSat} for a total duration of $\sim$135~ks. We found that the synchrotron convolved broken power-law model provided a better fit to the time averaged spectrum in the energy range 0.7--30 keV than the synchrotron convolved power-law  and logparabola models (see Table \ref{tab_Tavg}). 
In order to examine the spectral and flux properties over finer time bins, we used the method of time-resolved spectroscopy, where the total observation period ($\sim$135~ks) was divided into time segments of 5\,ksec. In each time bin, we found that the broken power-law model for the particle distribution  yielded a good fit statistic (see Table \ref{table:trs} ). 
The SXT, LAXPC,  UVIT-NUV and UVIT-FUV emission as measured by the \textit{AstroSat} instrument shows evidence for variability. 
In this work we mainly concentrated on interpreting the correlations between X-ray spectral parameters obtained from the time-resolved analysis.

 The lower energy particle index $\Gamma_1$ shows a marked anti-correlation with the 0.7--30 keV flux, the anti-correlation at X-ray energies between the photon index and flux as has been observed in other sources \citep[e.g. ][]{2006ApJ...646...61G,2015ApJ...807...79H,2017ApJ...848..103K}. However, since it is more straightforward to interpret a correlation between the particle number density at a given energy with its index, we estimate the number density using the normalisation of the fit. The  normalisation  parameter can be chosen to correspond to the number density at different reference energies. We found it optimal to choose an energy reference $E_{ref}\,(or\, \xi^2_{ref})= 1.21$\,keV which is smaller than the typical measured break energy of $E_{brk}\,(or\,\xi_{brk}^2)=2.86$\,keV, since the errors on estimated number density was least for such a reference energy. The particle index $\Gamma_1$ was again found to anti-correlate with the number density, but with a lower significance as compared to its correlation with flux. The normalised variation of the particle density $\delta n_\gamma/n_\gamma$ was found to be $\sim 0.1$ which is less than the observed variation in index $\Delta \Gamma \sim 0.5$. Moreover, if the index variation was shifted by $\sim 60$ ksec, a
correlation was seen between the index and number density, instead of the anti-correlation seen when such a shift was
not introduced.

We model the system as having an acceleration region where a power-law energy distribution of electrons is created and escape to a cooling region. In the cooling region, their distribution takes the form of a broken power-law with radiative cooling important for particles with energy greater than the break energy and not important for lower energies. We argue that any variability on time-scale longer than the time-scales of the system, would lead to a normalised variation of the  particle density at a reference energy, $\gamma_{ref} < \gamma_{brk}$ to be much larger than that of the index, $\Delta n_\gamma (\gamma_{ref})/n_\gamma (\gamma_{ref}) >> \Delta \Gamma_1$, which is contrary to the observations. On the other hand, if there is variability in the acceleration time-scale $\tau_{acc}$ and this variability itself has a time-scale of the same order as the acceleration one, then the observed amplitudes of $\Delta n_\gamma (\gamma_{ref})/n_\gamma (\gamma_{ref})$ and
$\Delta \Gamma_1$ can be explained. Moreover, such a model would predict a time-lag between the two and the data indeed shows that a near sinusoidal variation in the particle density is followed by a similar variation in index after a time-lag. This identification allows us to constrain physical parameters of the system such as the acceleration time scale $\tau_{acc} \sim  1.38\times 10^4 \delta/(1+z)$\,secs which for a Lorentz boosting factor $\delta \sim 15$ \citep{2018A&A...611A..44P}, turns out to be $1.97\times 10^5$\,secs. Then the escape time-scale can be estimated to be $\sim \tau_{acc}/(\Gamma_1-1) \sim 2.24\times 10^5$\,secs. If we assume that this corresponds to the $R/c$, then the size of the system turns out to be $R \sim 6.73\times 10^{15}$\,cms.

There are possible physical situations where the variation time-scale of the particle acceleration time-scale may be comparable to itself. A simplistic example would be where the acceleration time-scale for shock acceleration is expressed as  $nR_s/c$, where $R_s$ is the width of the shock and $n$ is the average number of times an electron has to cross the shock to gain substantial energy. Now if a variability is introduced due to the variation of the shock width, the time-scale of the variability would be $R_s/|\dot {R_s}|$, where $|\dot {R_s}|$ is the magnitude of the rate of change of the width. This variability time-scale would be of the same order as the acceleration one if $\dot {R_s} \sim c/n$.

We reemphasise that the observed time-lag $\sim 60$ ksecs is of the order of the length of the observation and the inference is based only one oscillation cycle.  What is required is significantly longer duration of observation (say 10 times the observation time used in the present work) to statistically confirm the the time-lag. 
Even for such an observation, complexities would arise if the magnetic field or the Doppler factor varies during the observation which would mean that the reference energy at which the particle density is being measured would not correspond to the same particle energy. It should also be noted that the broken power-law distribution used to fit the time-resolved spectra is at best an approximation to the real smooth distribution coming from a cooling region. The situation may be more complicated, if the micro-physics of the system is different, such as the escape or acceleration time-scale being dependent on the energy of the particle. Spectral curvature observed in several sources indicate that this indeed be the case  \citep{2004A&A...413..489M,2018MNRAS.478L.105J, 2018MNRAS.480.2046G, 2020MNRAS.492..796G, 2020MNRAS.tmp.2819G}.
Infact X-ray spectral modelling of 1ES 1959+650 spectra during the period 2015 August – 2017 November (when it was in a high flux state) required a  log-parabola model with a wide range of
curvature values (e.g. \cite{2016MNRAS.461L..26K, 2016MNRAS.457..704K, 2018MNRAS.473.2542K, 2018ApJS..238...13K, 2018A&A...611A..44P, 2019ApJ...885....8W}. This source also shows wide variation for the peak energy which ranges from
$\sim 0.1$ to $\sim 7.7$ keV \citep{2018ApJS..238...13K}. In summary, although the intrinsic spectral behaviour of blazars is expected to be complex, the relative amplitude of the particle density and index variations and the time-lag between them, can be effectively used to constrain the underlying physics.

\section{Acknowledgements}

This publication has made use of data from the \textit{Astrosat} mission of the Indian Space Research Organisation(ISRO),archived at the Indian Space Science Data Centre(ISSDC). This work has used the data from the Soft X-ray Telescope (SXT) developed at TIFR, Mumbai, and the SXT POC at TIFR is thanked for verifying and releasing the data via the ISSDC data archive and providing the necessary software tools. LaxpcSoft software is used for analysis of the LAXPC data and we acknowledge the LAXPC Payload Operation Center (TIFR, Mumbai). We thank the UVIT POC at IIA, Bangalore for their support. This research has  made  use  of  data,  software  and/or  web  tools  obtained  from the  High  Energy  Astrophysics  Science  Archive  Research  Center (HEASARC),  a  service  of  the  Astrophysics  Science  Division at NASA/GSFC and of the Smithsonian Astrophysical Observatory’s High Energy Astrophysics Division. 

\section{Data availability}
The data and the model used in this article will be shared on reasonable request to the corresponding author, Zahir Shah (email: zahir@iucaa.in) or Savithri H. Ezhikode (email: savithri@iucaa.in).
\bibliographystyle{mnras}
\bibliography{references} 

\bsp	
\label{lastpage}
\end{document}